\documentclass[cits]{PoS}
\usepackage{psfig,epsfig,colordvi,amssymb,amsmath,bbm}

\def\lsim{\mathrel{\rlap{\lower4pt\hbox{\hskip1pt$\sim$}}
    \raise1pt\hbox{$<$}}}                
\def\gsim{\mathrel{\rlap{\lower4pt\hbox{\hskip1pt$\sim$}}
    \raise1pt\hbox{$>$}}}                

\newcommand{\be}{\begin{equation}}
\newcommand{\ee}{\end{equation}}
\newcommand{\bea}{\begin{eqnarray}} 
\newcommand{\eea}{\end{eqnarray}}

\title{${\cal O}(a^2)$ corrections to 1-loop matrix elements of
  4-fermion operators with improved fermion/gluon actions}

\ShortTitle{${\cal O}(a^2)$ corrections to 1-loop matrix elements of
  4-fermion operators}

\author{Martha Constantinou$^a$\thanks{Work supported in part by the
    Research Promotion Foundation of Cyprus (Proposal Nr:
    TEXN/0308/17, $\rm ENI\Sigma X$/0506/17).} , Vittorio Lubicz$^b$, Haralambos
  Panagopoulos$^a$, Apostolos Skouroupathis$^{a\,*}$, \speaker{Fotos
    Stylianou}$^{a\,*}$ \\ 
 \llap{$^a$} Department of Physics, University of Cyprus\\
        P.O.Box 20537, Nicosia CY-1678, Cyprus\\ 
 \llap{$^b$} Dipartimento di Fisica, Universit\`a di Roma Tre and
 INFN, Sezione di Roma Tre,\\
        Via della Vasca Navale 84, I-00146 Roma, Italy\\ 
 E-mail: \email{marthac@ucy.ac.cy}, \email{lubicz@fis.uniroma3.it},
         \email{haris@ucy.ac.cy}, \email{php4as01@ucy.ac.cy},
         \email{fstyli01@ucy.ac.cy}}

\abstract{$\qquad$We calculate the corrections to the amputated Green's
  functions of 4-fermion operators, in 1-loop Lattice Perturbation
  theory. The novel aspect of our calculations is that they are
  carried out to second order in the lattice spacing, ${\cal O}(a^2)$.

  $\qquad$We employ the Wilson/clover action for massless fermions (also
  applicable for the twisted mass action in the chiral limit) and the
  Symanzik improved action for gluons. Our calculations have been
  carried out in a general covariant gauge. Results have been obtained
  for several popular choices of values for the Symanzik coefficients
  (Plaquette, Tree-level Symanzik, Iwasaki, TILW and DBW2 action).

  $\qquad$We pay particular attention to $\Delta F=2$ operators, both
  Parity Conserving and Parity Violating ($F$ stands for flavour: $S,\,C,\,B$). 
  We study the mixing
  pattern of these operators, to ${\cal O}(a^2)$, using the
  appropriate projectors. Our results for the corresponding
  renormalization matrices are given as a function of a large number of
  parameters: coupling constant, clover parameter, number of colors,
  lattice spacing, external momentum and gauge parameter.

  $\qquad$The ${\cal O}(a^2)$ correction terms (along with our previous 
  ${\cal O}(a^2)$ calculation of $Z_\Psi$) are essential
  ingredients for minimizing the lattice artifacts which are present
  in non-perturbative evaluations of renormalization constants with
  the RI$'$-MOM method.
 
\bigskip
  A longer write-up of this work, including non-perturbative results,
  is in preparation together with members of the ETM Collaboration
\cite{CLPSSETMC}. }


\FullConference{The XXVII International Symposium on Lattice Field Theory\\
         July 26-31, 2009\\
         Peking University, Beijing, China}

\begin{document}

\section{Introduction}
A number of flavour-changing processes are currently under study in
Lattice simulations. Among the most common examples are the
decay $K \rightarrow \pi \pi$ and $K^0$--$\bar K^0$ oscillations. 
From experimental evidence, we know that these
weak processes violate the CP symmetry. In theory, the calculation of
the amount of CP violation in $K^0$--$\bar K^0$ oscillations
requires the knowledge of $B_K$.

The Kaon $B_K$ parameter is obtained from the $\Delta S = 2$ weak
matrix element:
\be
B_K = \frac{\langle \bar K^0|\hat O^{\Delta S=2}| K^0 \rangle}
           {\frac{8}{3}\langle \bar K^0|\bar s \gamma_\mu d|0 \rangle \, \langle 0|\bar s \gamma_\mu d|K^0 \rangle} \, ,
\ee
where $s$ and $d$ stand for strange and down quarks, and $\hat
O^{\Delta S=2}$ is the effective 4-quark interaction renormalized
operator, corresponding to the bare operator:
\be
O^{\Delta S=2} = (\bar s \gamma_\mu^{\rm L} d)(\bar s \gamma_\mu^{\rm L} d) ,
\qquad \gamma_\mu^{\rm L}=\gamma_\mu(\mathbbm{1}-\gamma_5).
\ee
The above operator splits into parity-even and parity-odd parts; in
standard notation:
$O^{\Delta S=2} = O_{VV+AA}^{\Delta S=2} - O_{VA+AV}^{\Delta S=2}$. 
Since the above weak process is
simulated in the framework of Lattice QCD, where Parity is a symmetry,
the parity-odd part gives no contribution to the $K^0$--$\bar K^0$
matrix element. Thus, we conclude
that $B_K$ can be extracted from the correlator ($x_0\!>\!0$,
$y_0\!<\!0$): 
{\small{
\be 
C_{KOK}(x,y) = \langle (\bar{d}\gamma_5 s)(x) \hat{O}_{VV+AA}^{\Delta S=2}(0) (\bar{d}\gamma_5 s)(y) \rangle, \qquad
O_{VV+AA}^{\Delta S=2} = (\bar{s} \gamma_\mu d) (\bar{s}
\gamma_\mu d) + (\bar{s} \gamma_\mu\gamma_5 d) (\bar{s}
\gamma_\mu\gamma_5 d) \, ,
\label{VVAA}
\ee
}}
\hspace{-0.15cm}where $O_{VV+AA}^{\Delta S=2}$ is the bare operator and
$\hat{O}_{VV+AA}^{\Delta S=2}$ is the respective renormalized
operator.

In place of the operator in Eq. (\ref{VVAA}) it is advantageous to
use a four-quark 
operator with a different flavour content ($s$, $d$, $s'$, $d'$), and
with $\Delta S=\Delta s + \Delta s'=2$,
namely \cite{FR}: 
\be
 {\cal O}_{VV+AA}^{\Delta S=2} =
(\bar{s} \gamma_\mu d) (\bar{s}' \gamma_\mu d')
+ (\bar{s} \gamma_\mu\gamma_5 d) (\bar{s}' \gamma_\mu\gamma_5 d')
+ (\bar{s} \gamma_\mu d') (\bar{s}' \gamma_\mu d)
+ (\bar{s} \gamma_\mu\gamma_5 d') (\bar{s}' \gamma_\mu\gamma_5 d) \, ,
\label{VVAAprime}
\ee
where now the correlator is given by:
%
$C_{K{\cal O}K'}(x,y) = \langle
(\bar{d}\gamma_5 s)(x) \, 2 {\cal O}_{VV+AA}^{\Delta S=2}(0)
(\bar{d}'\gamma_5 s')(y) \rangle $. 
%
Making use of Wick's theorem one checks the equality:
$C_{K{\cal O}K'}(x,y) = C_{KOK}(x,y)$,
which means that both correlators contain the same physical
information.

The aforementioned matrix elements are very sensitive to various systematic
errors. A major issue facing Lattice Gauge Theory, since its early
days, has been the reduction of effects induced by the finiteness of
lattice spacing $a$, in order to better approach the elusive continuum
limit. 

In order to obtain reliable non-perturbative estimates of physical
quantities (i.e. improving the accuracy of $B_K$) it is essential to
keep under control the ${\cal O}(a)$ systematic errors in simulations
or, additionally, reduce the lattice artifacts in numerical results. 
Such a reduction, regarding renormalization functions, can be achieved by
subtracting appropriately the ${\cal O}(a^2)$ perturbative correction terms
presented in this paper, from respective non-perturbative results.

In this paper we calculate the amputated Green's functions and the
renormalization matrices of the complete basis of 20 four-fermion
operators of dimension six which
do not need power subtractions (i.e. mixing occurs only with other
operators of equal dimensions). The calculations are carried out up to
1-loop in Lattice Perturbation theory and up to ${\cal O}(a^2)$ in
lattice spacing. Our results are immediately applicable to other
$\Delta F=2$ processes of great phenomenological interest, such as
$D-\bar D$ or $B-\bar B$ mixing. Let us also mention that in generic
new physics models (i.e. beyond the standard model), the complete
basis of 4-fermion operators contributes to neutral meson mixing
amplitudes; this is the case for instance of SUSY models (see
e.g. \cite{SUSY}).
\newpage

\section{Amputated Green's functions of 4-fermion $\Delta S= \Delta s + \Delta s'=2$ operators.}

In this work we evaluate, up to ${\cal O}(a^2)$, the 1-loop matrix
element of the 4-fermion operators\footnote{The superscript letter F
  stands for Fierz.}: 
{\small{
\bea
{\cal O}_{XY} \equiv (\bar{s}\,X\,d)(\bar{s}'\,Y\,d')\equiv 
\sum_x \sum_{c,d} \sum_{k_1,\,k_2,\,k_3,\,k_4}
\Bigl(\bar{s}_{{k_1}}^c(x)  \,X_{{k_1 k_2}}\,  d_{{k_2}}^c(x)\Bigr) 
\Bigl(\bar{s'}_{{k_3}}^d(x) \,Y_{{k_3 k_4}}\, {d'}_{{k_4}}^d(x)\Bigr) \\
\label{O_XY}
{\cal O}^F_{XY} \equiv (\bar{s}\,X\,d')(\bar{s}'\,Y\,d)\equiv 
\sum_x \sum_{c,d} \sum_{k_1,\,k_2,\,k_3,\,k_4}
\Bigl(\bar{s}_{{k_1}}^c(x)  \,X_{{k_1 k_2}}\, {d'}_{{k_2}}^c(x)\Bigr) 
\Bigl(\bar{s'}_{{k_3}}^d(x) \,Y_{{k_3 k_4}}\, d_{{k_4}}^d(x)\Bigr)
\label{O^F_XY}
\eea
}}
\hspace{-0.15cm}with a generic initial state: 
$\bar{d'}_{{i_4}}^{{a_4}}(p_4)\,{s'}_{{i_3}}^{{a_3}}(p_3) |0\rangle $,
and a generic final state:
$\langle 0 | \bar{d}_{{i_2}}^{{a_2}}(p_2)\,s_{{i_1}}^{{a_1}}(p_1)$.
Spin indices are denoted by $i,\,k$, and color indices by
$a,\,c,\,d$, while  
$X$ and $Y$ correspond to the following set of products of the Dirac
matrices:
{\small{
\be
X,\,Y = \{\mathbbm{1},\, \gamma^5,\, \gamma_\mu,\, \gamma_\mu \gamma^5,
  \,\sigma_{\mu\nu}, \,\gamma^5\sigma_{\mu\nu}\} \equiv
  \{S,P,V,A,T,\tilde T \}; 
  \qquad \sigma_{\mu\nu}=\frac{1}{2}[\gamma_\mu,\gamma_\nu].
\label{Gamma}
\ee
}}
\hspace{0.65cm}Our calculations are performed using massless fermions described by
the Wilson/clover action. By taking $m_f=0$, our results 
are identical also for the twisted mass action and the
Osterwalder-Seiler action in the chiral limit (in the so called
twisted mass basis). For gluons we employ a 3-parameter family of
Symanzik improved actions, which comprises all common gluon actions
(Plaquette, tree-level Symanzik, Iwasaki, DBW2, L\"uscher-Weisz). 
Conventions and notations for the actions used, as well as algebraic
manipulations involving the evaluation of 1-loop Feynman diagrams (up
to ${\cal O}(a^2)$), are described in detail in Ref. \cite{Propolemiko}.

To establish notation and normalization, let us first write the
tree-level expression for the amputated Green's functions of the
operators ${\cal O}_{XY}$ and ${\cal O}^F_{XY}$:
{\small{
\be
\Lambda^{XY}_{tree}(p_1,p_2,p_3,p_4,r_s,r_d,r_{s'},r_{d'})_
{{i_1 i_2 i_3 i_4}}^{{a_1 a_2 a_3 a_4}} = 
X_{i_1 i_2}\, Y_{i_3 i_4}\, \delta_{a_1 a_2}\,\delta_{a_3 a_4} ,
\label{L_XY_tree}
\ee
\be
(\Lambda^F)^{XY}_{tree}(p_1,p_2,p_3,p_4,r_s,r_d,r_{s'},r_{d'})_
{{i_1 i_2 i_3 i_4}}^{{a_1 a_2 a_3 a_4}} = 
- X_{i_1 i_4}\, Y_{i_3 i_2}\, \delta_{a_1 a_4}\,\delta_{a_3 a_2} ,
\ee
}}
\hspace{-0.15cm}where $r$ is the Wilson parameter, one for each flavour. 

\begin{figure}
\includegraphics[height=3.20truecm]{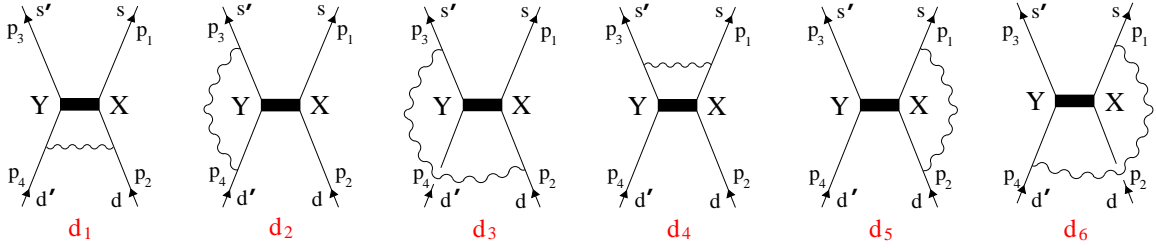}
\caption{1-loop diagrams contributing to the amputated Green's
function of the 4-fermi operator ${\cal O}^{XY}$. Wavy (solid) lines
represent gluons (fermions).}
\label{fig1}
\end{figure}

We continue with the first quantum corrections. There are twelve 1-loop
diagrams that enter our 4-fermion calculation, six for each operator
${\cal O}_{XY}$, ${\cal O}^F_{XY}$. The diagrams $d_1-d_6$
corresponding to the operator ${\cal O}_{XY}$ are illustrated in
Fig. \ref{fig1}. The other six diagrams,
$d^F_1-d^F_6$, involved in the Green's function of ${\cal O}^F_{XY}$ 
are similar to $d_1-d_6$, and may be obtained from $d_1-d_6$ by
interchanging the fermionic fields $d$ and $d'$ along with their
momenta, color and spin indices, and respective Wilson parameters. 
%
%
%

The only diagrams that need to be calculated from first principles are
$d_1$, $d_2$ and $d_3$, while the rest can be expressed in terms of the
first three. In particular, the expressions for the amputated Green's
functions $\Lambda_{d_4}^{XY}-\Lambda_{d_6}^{XY}$ can be obtained via the
following relations:
{\small{
\bea
\Lambda_{d_4}^{XY}(p_1,p_2,p_3,p_4,r_s,r_d,r_{s'},r_{d'})_{i_1 i_2 i_3 i_4}^{a_1 a_2 a_3 a_4} &=&
\left(\Lambda_{d_1}^{XY}({-}p_2,{-}p_1,{-}p_4,{-}p_3,r_d,r_s,r_{d'},r_{s'})_{i_2 i_1 i_4 i_3}^{a_2 a_1 a_4 a_3}\right)^\star \label{Lambda_4},\\ 
\Lambda_{d_5}^{XY}(p_1,p_2,p_3,p_4,r_s,r_d,r_{s'},r_{d'})_{i_1 i_2 i_3 i_4}^{a_1 a_2 a_3 a_4} &=&\phantom{\Bigl(}
\Lambda_{d_2}^{YX}      (p_3,p_4,p_1,p_2,r_{s'},r_{d'},r_s,r_d)_{i_3 i_4 i_1 i_2}^{a_3 a_4 a_1 a_2},\\ 
\Lambda_{d_6}^{XY}(p_1,p_2,p_3,p_4,r_s,r_d,r_{s'},r_{d'})_{i_1 i_2 i_3 i_4}^{a_1 a_2 a_3 a_4} &=&\phantom{\Bigl(}
\Lambda_{d_3}^{YX} (p_3,p_4,p_1,p_2,r_{s'},r_{d'},r_s,r_d)_{i_3 i_4 i_1 i_2}^{a_3 a_4 a_1 a_2}.
\eea
}}
\hspace{-0.15cm} Once we have constructed
$\Lambda_{d_4}^{XY}-\Lambda_{d_6}^{XY}$ we can use relation:
{\small{
\be
(\Lambda^F)_{d_j}^{XY}(p_1,p_2,p_3,p_4,r_s,r_d,r_{s'},r_{d'})_{i_1 i_2 i_3 i_4}^{a_1 a_2 a_3 a_4} =
-\Lambda_{d_j}^{XY}      (p_1,p_4,p_3,p_2,r_s,r_{d'},r_{s'},r_d)_{i_1 i_4 i_3 i_2}^{a_1 a_4 a_3 a_2}, 
\ee
}}
\hspace{-0.15cm}to derive the expressions for $(\Lambda^F)_{d_1}^{XY}-(\Lambda^F)_{d_6}^{XY}$.
From the amputated Green's functions for all
twelve diagrams we can write down the total 1-loop
expressions for the operators ${\cal O}_{XY}$ and ${\cal O}^F_{XY}$:
{\small{
\be
\Lambda_{1-loop}^{XY} = \sum_{j=1}^{6}\Lambda_{d_j}^{XY},\qquad
(\Lambda^F)_{1-loop}^{XY}= \sum_{j=1}^{6}(\Lambda^F)_{d_j}^{XY}.
\label{L^F_XY_1loop}
\ee
}}

In our algebraic expressions for the 1-loop amputated Green's
functions $\Lambda_{d_1}^{XY}$, $\Lambda_{d_2}^{XY}$ and $\Lambda_{d_3}^{XY}$
we kept the Wilson parameters for each quark field distinct, that is:
$\,r_s$, $r_d$, $r_{s'}$, $r_{d'}$ for the quark fields $s$, $d$, $s'$ 
and $d'$ respectively. For the required numerical integration of the
algebraic expressions of the integrands, corresponding to each Feynman
diagram, we are forced to
choose the square of the value for each $r$ parameter. As in all
present day simulations, we set:
\be
r_s^2=r_d^2=r_{s'}^2=r_{d'}^2 \equiv 1. 
\ee
Concerning the external momenta $p_i$ (shown explicitly in
Fig. \ref{fig1}) we have chosen to evaluate the
amputated Green's functions at the renormalization point:
\be
p_1 = p_2 = p_3 = p_4 \equiv p.
\ee
It is easy and not time consuming to repeat the calculations for other
choices of Wilson parameters and for other renormalization prescriptions.
The final 1-loop expressions for $\Lambda_{d_1}^{XY}$, $\Lambda_{d_2}^{XY}$
and $\Lambda_{d_3}^{XY}$, up to ${\cal O}(a^2)$, are obtained as a
function of: the coupling constant $g$, clover
parameter $c_{SW}$, number of colors $N_c$, lattice spacing $a$, external
momentum $p$ and gauge parameter $\lambda$.

The crucial point of our calculation is the correct extraction of the
full ${\cal O}(a^2)$ dependence from loop integrands with strong IR
divergences (convergent only beyond 6 dimensions). The singularities
are isolated using the procedure explained in
Ref. \cite{Propolemiko}. In order to reduce the number of strong IR
divergent integrals, appearing in diagram $d_1$, we have inserted the
identity below into selected 3-point functions: 
{\small{
\be
1=\frac{1}{\widehat{a\,p}^2}\Bigl(\widehat{k+a\,p}^2 +
\widehat{k-a\,p}^2 - 2 \hat k^2 + 16 \sum_\sigma \sin(k_\sigma)^2
\sin(a p_\sigma)^2\Bigr),
\label{identity} 
\ee
}}
\hspace{-0.15cm}where $\hat{q}^2=4\sum_\mu \sin^2(\frac{q_\mu}{2})$ and $k\,(p)$ is
the loop (external) momentum. The common factor in Eq. (\ref{identity})
can be treated by Taylor expansion. For our calculations it was
necessary only to ${\cal O}(a^0)$: 
{\small{
\be
\frac{1}{\widehat{a\,p}^2} =
\frac{1}{a^2\,p^2} + \frac{\sum_\sigma p_\sigma^4}{(p^2)^2} +{\cal O}(a^2\,p^2) .
\ee
}}
\hspace{-0.15cm}Here we present one of the four integrals
with strong IR divergences that enter in this calculation:

{\small{
\bea
\hspace{-0.15cm}&&\hspace{-0.15cm}\Blue{\int_{-\pi}^\pi
\frac{d^4k}{(2\pi)^4}\frac{\sin(k_\mu)\,\sin(k_\nu)}{\hat k^2\, 
\widehat{k+a\,p}^2\, \widehat{k-a\,p}^2}} = 
\delta_{\mu\nu} \Bigl[\,0.002457072288 - \frac{\ln(\Red{a^2}p^2)}{64\pi^2}
+\Red{a^2}\,p^2 \Bigl( 0.00055270353(6) - \frac{\ln(\Red{a^2}p^2)}{512\pi^2}\Bigr)\nonumber\\
\hspace{-0.15cm}&&\hspace{-0.15cm}- \Red{a^2}\,p_\mu^2 \Bigl(0.0001282022(1)
  +\frac{\ln(\Red{a^2}p^2)}{768\pi^2}\Bigr)
+0.000157122310\,\Red{a^2}\frac{\sum_\sigma p_\sigma^4}{p^2}\Bigl]\,
+ \Red{a^2}\,p_\mu\,p_\nu \Bigl[0.001870841540 \frac{1}{\Red{a^2}\,p^2}\nonumber\\
\hspace{-0.15cm}&&\hspace{-0.15cm}-0.00029731225(4)+\frac{\ln(\Red{a^2}p^2)}{768\pi^2} 
-0.000047949674 \frac{(p_\mu^2+p_\nu^2)}{p^2}
+0.000268598599\frac{\sum_\sigma p_\sigma^4}{(p^2)^2} \Bigl] +{\cal O}(\Red{a^4}\,p^4).\nonumber
\eea
}} 

\noindent The results for the other three integrals can be found in Ref. \cite{Propolemiko}.
Integrands with simple IR divergences (convergent beyond 4 dimensions)
can be handled by well-known techniques. 

Due to lack of space we present only the results for
$\Lambda_{d_1}^{XY}$ and for the special choices: 
$c_{SW}=0$, $\lambda=0$ (Landau Gauge), $r_s=r_d=r_{s'}=r_{d'}=1$, and
tree-level Symanzik action: 
%
{\small{
\bea
&&\Blue{\Lambda_{d_1}^{XY}(p)_{i_1\,i_2\,i_3\,i_4}^{a_1\,a_2\,a_3\,a_4}}=\frac{g^2}{16\pi^2}
\left(\delta_{a_1\,a_4}\delta_{a_3\,a_2}-\frac{\delta_{a_1\,a_2}\delta_{a_3\,a_4}}{N_c}\right)
\times
\bigg \{
X_{i_1\,i_2}Y_{i_3\,i_4}\left[-\frac{1}{2}\ln(\Red{a^2}p^2)-0.05294139(2)\right]\nonumber \\
&&+\sum_\mu (X\gamma^\mu)_{i_1\,i_2}(Y\gamma^\mu)_{i_3\,i_4}\left[-0.507914049(6)\right] 
+ \sum_{\mu,\nu} (X\gamma^\mu\gamma^\nu)_{i_1\,i_2}(Y\gamma^\mu\gamma^\nu)_{i_3\,i_4}\left[\frac{1}{8}\ln(\Red{a^2}p^2)+ 0.0185984988(9)\right] \nonumber\\
&&+ \sum_{\mu,\nu,\rho} (X\gamma^\mu\gamma^\rho)_{i_1\,i_2}(Y\gamma^\nu\gamma^\rho)_{i_3\,i_4}\left[0.3977157268533\frac{p_\mu p_\nu}{p^2}\right]+\Red{a}\Blue{(\Lambda_{{\cal O}(a^1)})^{XY}_{d_1}}+\Red{a^2}\Blue{(\Lambda_{{\cal O}(a^2)})^{XY}_{d_1}} 
\bigg \} ,
\eea
}}
\hspace{-0.15cm}where:
{\small{
\bea
&&\hspace{-1.0cm} \Blue{(\Lambda_{{\cal O}(a^1)})^{XY}_{d_1}} = \sum_\mu\Big( (X\gamma^\mu)_{i_1\,i_2} \,Y_{i_3\,i_4}+X_{i_1\,i_2}(Y\gamma^\mu)_{i_3\,i_4}\Big)\times\left[ip_\mu\left(-\frac{1}{4}\ln(\Red{a^2}p^2)+ 0.09460083(1)\right)\right]\nonumber \\
&&\hspace{-1.0cm} + \sum_{\mu,\nu}\Big( (X\gamma^\mu\gamma^\nu)_{i_1\,i_2} (Y_{i_3\,i_4}\gamma^\nu)+(X\gamma^\nu)_{i_1\,i_2}(Y\gamma^\mu\gamma^\nu)_{i_3\,i_4}\Big)\times\left[ip_\mu\left(\frac{1}{16}\ln(\Red{a^2}p^2)+ 0.1692905881(6)\right)\right],
\eea
}}
\hspace{-0.15cm}and:
{\small{
\bea
&&\hspace{-1.3cm}\Blue{(\Lambda_{{\cal O}(a^2)})^{XY}_{d_1}} = X_{i_1\,i_2}Y_{i_3\,i_4}\left[p^2\left(-\frac{17}{72}\,\ln(\Red{a^2}p^2)+
1.32362250(4)\right)+0.06213649(4)\frac{\sum_\sigma p_\sigma^4}{p^2} \right]\nonumber \\
&&\hspace{-1.3cm} + \sum_\mu (X\gamma^\mu)_{i_1\,i_2}(Y\gamma^\mu)_{i_3\,i_4}\left[p^2\left(-\frac{7}{48}\ln(\Red{a^2}p^2) + 0.059895142(8)\right)+ 1.01694823(2) p_\mu^2 \right] \nonumber \\
&&\hspace{-1.3cm} + \sum_{\mu,\nu}\Big((X\gamma^\mu\gamma^\nu)_{i_1\,i_2}Y_{i_3\,i_4}+X_{i_1\,i_2}(Y\gamma^\mu\gamma^\nu)_{i_3\,i_4}\Big)\times \left[0.00592406(2) \frac{p_\nu p_\mu^3}{p^2}\right] \nonumber \\
&&\hspace{-1.3cm} + \sum_{\mu,\nu} (X\gamma^\mu)_{i_1\,i_2} (Y\gamma^\nu)_{i_3\,i_4} \left[p_\mu p_\nu \left(-\frac{1}{6}\ln(\Red{a^2}p^2)-0.19915360(1)\right)\right] \nonumber \\
&&\hspace{-1.3cm} + \sum_{\mu,\nu} (X\gamma^\mu\gamma^\nu)_{i_1\,i_2} (Y\gamma^\mu\gamma^\nu)_{i_3\,i_4}\bigg[p^2\left(\frac{7}{240}\,\ln(\Red{a^2}p^2)-0.089628048(6)\right)-0.048180850\frac{\sum_\sigma p_\sigma^4}{p^2}\nonumber \\
&&\hspace{-1.3cm} \hspace{4.25cm} + p_\mu^2\left(-\frac{29}{180}\,\ln(\Red{a^2}p^2)+0.16608907(3)\right)\bigg] \nonumber \\
&&\hspace{-1.3cm} + \sum_{\mu,\nu,\rho} (X\gamma^\mu\gamma^\rho)_{i_1\,i_2} (Y\gamma^\nu\gamma^\rho)_{i_3\,i_4}\bigg[p_\mu p_\nu\left(\frac{41}{360}\,\ln(\Red{a^2}p^2)-0.21865900(2)+0.140961390\frac{\sum_\sigma p_\sigma^4}{p^2}\right)\nonumber \\
&&\hspace{-1.3cm} \hspace{4.45cm} - 0.110138790\frac{(p_\mu^3 p_\nu + p_\mu p_\nu^3)}{p^2}-0.477634781(8)\frac{p_\mu p_\nu p_\rho^2}{p^2}\bigg] .
\eea
}}
%
\hspace{-0.15cm}Similar expressions exist for
$\Lambda_{d_2}^{XY}$ and $\Lambda_{d_3}^{XY}$.

\newpage
\section{Mixing and Renormalization of ${\cal O}_{XY}$ and ${\cal O}^F_{XY}$ on the lattice.}

The matrix element $\langle \bar K^0|O^{\Delta S=2}_{VV+AA}| K^0 \rangle$ 
is very sensitive to various systematic errors. The main roots of this
problem are: 
{\bf a)} ${\cal O}(a)$ systematic errors due to numerical integration,
{\bf b)} the operator $O^{\Delta S=2}_{VV+AA}$ mixes with other 4-fermion
$\Delta S=2$ operators of dimension six. Mixing with operators of lower
dimensionality is impossible because there is no candidate
$\Delta S=2$ operator.

In order to address these problems we have calculated the mixing
pattern (renormalization matrices) of the Parity Conserving and Parity
Violating 4-fermion $\Delta S=2$ operators (defined below), by using
the amputated Green's functions obtained in the previous
section. A more extensive theoretical background and non-perturbative
results, concerning renormalization matrices of 4-fermion operators,
can be found in Ref. \cite{DGMTV} (see also
\cite{FR,BGLMPR,MPSTV}). Next we summarize all important relations
from Ref. \cite{DGMTV} needed for the present calculation.

One can construct a complete basis of 20
independent operators which have the symmetries of the generic QCD Wilson lattice
action (Parity $P$, Charge
conjugation $C$, Flavour exchange symmetry $S {\equiv} (d
\leftrightarrow d')$, Flavour Switching symmetries $S' {\equiv}
(s \leftrightarrow d , s' \leftrightarrow d')$ and
$S'' {\equiv} (s \leftrightarrow d' , d \leftrightarrow s')$), with 4 degenerate quarks.
This basis can be decomposed into smaller independent bases according
to the discrete symmetries $P,\,S,\,CPS',\,CPS''$. Following the notation
of Ref. \cite{DGMTV} we have 10 Parity Conserving operators, $Q$,
($P{=}+1,\,S{=}\pm 1$)
and 10 Parity Violating operators, $\cal Q$, ($P{=}-1,\,S{=}\pm 1$):

\noindent
\begin{minipage}{8cm}
\bea
\begin{cases}
Q_1^{S=\pm 1}\equiv \frac{1}{2}\left[{\cal O}_{VV} \pm {\cal O}^F_{VV}\right]+\frac{1}{2}\left[{\cal O}_{AA} \pm {\cal O}^F_{AA}\right],\nonumber\\[0.4ex]
Q_2^{S=\pm 1}\equiv \frac{1}{2}\left[{\cal O}_{VV} \pm {\cal O}^F_{VV}\right]-\frac{1}{2}\left[{\cal O}_{AA} \pm {\cal O}^F_{AA}\right],\nonumber\\[0.4ex]
Q_3^{S=\pm 1}\equiv \frac{1}{2}\left[{\cal O}_{SS} \pm {\cal O}^F_{SS}\right]-\frac{1}{2}\left[{\cal O}_{PP} \pm {\cal O}^F_{PP}\right],\nonumber\\[0.4ex]
Q_4^{S=\pm 1}\equiv \frac{1}{2}\left[{\cal O}_{SS} \pm {\cal O}^F_{SS}\right]+\frac{1}{2}\left[{\cal O}_{PP} \pm {\cal O}^F_{PP}\right],\nonumber\\[0.4ex]
Q_5^{S=\pm 1}\equiv \frac{1}{2}\left[{\cal O}_{TT} \pm {\cal O}^F_{TT}\right],\nonumber
\end{cases}
\eea
\end{minipage}
\hfill
\begin{minipage}{7.5cm}
\bea
&\begin{cases}
{\cal Q}_1^{S=\pm 1}\equiv \frac{1}{2}\left[{\cal O}_{VA} \pm {\cal O}^F_{VA}\right]+\frac{1}{2}\left[{\cal O}_{AV} \pm {\cal O}^F_{AV}\right],\nonumber
\end{cases}\\
&\begin{cases}
{\cal Q}_2^{S=\pm 1}\equiv \frac{1}{2}\left[{\cal O}_{VA} \pm {\cal O}^F_{VA}\right]-\frac{1}{2}\left[{\cal O}_{AV} \pm {\cal O}^F_{AV}\right],\nonumber\\
{\cal Q}_3^{S=\pm 1}\equiv \frac{1}{2}\left[{\cal O}_{PS} \pm {\cal O}^F_{PS}\right]-\frac{1}{2}\left[{\cal O}_{SP} \pm {\cal O}^F_{SP}\right],\nonumber
\end{cases}\\
&\begin{cases}
{\cal Q}_4^{S=\pm 1}\equiv \frac{1}{2}\left[{\cal O}_{PS} \pm {\cal O}^F_{PS}\right]+\frac{1}{2}\left[{\cal O}_{SP} \pm {\cal O}^F_{SP}\right],\nonumber\\
{\cal Q}_5^{S=\pm 1}\equiv \frac{1}{2}\left[{\cal O}_{T\tilde T} \pm {\cal O}^F_{T\tilde T}\right].\nonumber
\end{cases}
\eea
\end{minipage}
%

\phantom{-------}
\noindent Summation over all independent Lorentz indices (if any),
of the Dirac matrices, is implied. The operators shown above are
grouped together according to their mixing pattern. This implies that
the renormalization matrices $Z^{S=\pm 1}$ (${\cal Z}^{S=\pm 1}$), for
the Parity Conserving (Violating) operators, have the form:
{\small{
\be
Z^{S=\pm 1}
=
\left(\begin{array}{rrrrr}
Z_{11}\,\, & Z_{12}\,\, & Z_{13}\,\, & Z_{14}\,\, & Z_{15} \\
Z_{21}\,\, & Z_{22}\,\, & Z_{23}\,\, & Z_{24}\,\, & Z_{25} \\
Z_{31}\,\, & Z_{32}\,\, & Z_{33}\,\, & Z_{34}\,\, & Z_{35} \\
Z_{41}\,\, & Z_{42}\,\, & Z_{43}\,\, & Z_{44}\,\, & Z_{45} \\
Z_{51}\,\, & Z_{52}\,\, & Z_{53}\,\, & Z_{54}\,\, & Z_{55} 
\end{array}\right)^{S=\pm 1},
\quad
{\cal Z}^{S=\pm 1}
=
\left(\begin{array}{rrrrr}
 {\cal Z}_{11}  &0\,\,         &0\,\,         &0\,\,        &0\,\,  \\
 0\,\,         &{\cal Z}_{22}  &{\cal Z}_{23}  &0\,\,        &0\,\,  \\
 0\,\,         &{\cal Z}_{32}  &{\cal Z}_{33}  &0\,\,        &0\,\,  \\
 0\,\,         &0\,\,         &0\,\,         &{\cal Z}_{44}  &{\cal Z}_{45} \\
 0\,\,         &0\,\,         &0\,\,         &{\cal Z}_{54}  &{\cal Z}_{55}
\end{array}\right)^{S=\pm 1}.
\ee
}}

Now the renormalized Parity Conserving (Violating) operators,
$\hat{Q}^{S=\pm 1}$ ($\hat{\cal Q}^{S=\pm 1}$), are defined via the
equations:
\be
{\hat Q}_l^{S=\pm 1} = Z^{S=\pm 1}_{lm} \cdot Q^{S=\pm 1}_{m} ,\quad
\hat{\cal Q}^{S=\pm 1}_l = {\cal Z}^{S=\pm 1}_{lm} \cdot {\cal Q}^{S=\pm 1}_m,
\ee
where $l,m = 1,\dots ,5$ (a sum over $m$ is implied).
The renormalized amputated Green's functions $\hat{L}^{S=\pm 1}$
($\hat{\cal L}^{S=\pm 1}$) corresponding to $Q^{S=\pm 1}$ (${\cal Q}^{S=\pm 1}$), are given
in terms of their bare counterparts $L^{S=\pm 1}$ (${\cal L}^{S=\pm 1}$) through:
\be
{\hat L}_l^{S=\pm 1} = Z_\Psi^{-2} Z^{S=\pm 1}_{lm} \cdot L^{S=\pm 1}_{m},\quad
\hat{\cal L}^{S=\pm 1}_l = Z_\Psi^{-2} {\cal Z}^{S=\pm 1}_{lm} \cdot {\cal L}^{S=\pm 1}_{m},
\label{rengreen}
\ee
where $Z_\Psi$ is the quark field renormalization constant.

In order to calculate the renormalization matrices $Z^{S=\pm 1}$
(${\cal Z}^{S=\pm 1}$), we make use of the appropriate Parity
Conserving (Violating) Projectors $P^{S=\pm 1}$ (${\cal P}^{S=\pm 1}$): 

\begin{minipage}{6cm}
\bea
P_1^{S=\pm 1}&\equiv& + \frac{\Pi_{VV}+\Pi_{AA}}{64N_c(N_c\pm 1)},\nonumber\\
P_2^{S=\pm 1}&\equiv& + \frac{\Pi_{VV}-\Pi_{AA}}{64(N_c^2-1)} 
               \pm \frac{\Pi_{SS}-\Pi_{PP}}{32N_c(N_c^2-1)} ,\nonumber\\
P_3^{S=\pm 1}&\equiv&  \pm \frac{\Pi_{VV}-\Pi_{AA}}{32N_c(N_c^2-1)}
               +  \frac{\Pi_{SS}-\Pi_{PP}}{16(N_c^2-1)},\nonumber\\
P_4^{S=\pm 1}&\equiv& + \frac{\Pi_{SS}+\Pi_{PP}}{\frac{32N_c(N_c^2-1)}{2N_c\pm 1}}
                \mp \frac{\Pi_{TT}}{32N_c(N_c^2-1)},\nonumber\\
P_5^{S=\pm 1}&\equiv& \mp \frac{\Pi_{SS}+\Pi_{PP}}{32N_c(N_c^2-1)}
               + \ \frac{\Pi_{TT}}{\frac{96N_c(N_c^2-1)}{2N_c\mp 1}} ,\nonumber
\eea
\end{minipage}
\hfill
\begin{minipage}{7.5cm}
\bea
{\cal P}_1^{S=\pm 1}&\equiv& - \frac{\Pi_{VA}+\Pi_{AV}}{64N_c(N_c\pm 1)},\nonumber\\
{\cal P}_2^{S=\pm 1}&\equiv& - \frac{\Pi_{VA}-\Pi_{AV}}{64(N_c^2-1)}
                 \mp \frac{\Pi_{SP}-\Pi_{PS}}{32N_c(N_c^2-1)},\nonumber\\
{\cal P}_3^{S=\pm 1}&\equiv& \mp \frac{\Pi_{VA}-\Pi_{AV}}{32N_c(N_c^2-1)}
                    -  \frac{\Pi_{SP}-\Pi_{PS}}{16(N_c^2-1)},\nonumber\\
{\cal P}_4^{S=\pm 1}&\equiv& + \frac{\Pi_{SP}+\Pi_{PS}}{\frac{32N_c(N_c^2-1)}{2N_c\pm 1}}
                 \mp \frac{\Pi_{T\tilde T}}{32N_c(N_c^2-1)} ,\nonumber\\
{\cal P}_5^{S=\pm 1}&\equiv& \mp \frac{\Pi_{SP}+\Pi_{PS}}{32N_c(N_c^2-1)}
                   + \frac{\Pi_{T\tilde T}}{\frac{96N_c(N_c^2-1)}{2N_c\mp 1}} ,\nonumber
\eea
\end{minipage}
%

\phantom{-------}
\noindent where $\Pi_{XY}\equiv (X_{i_2 i_1} \otimes Y_{i_4 i_3})\delta_{a_2 a_1}\delta_{a_4 a_3}$. 
Again, summation is implied over all independent Lorentz indices (if any) of the
Dirac matrices. The above Projectors are chosen to obey
the following orthogonality conditions:
\be
Tr(P_l^{S=\pm 1} \cdot L_{m\,(tree)}^{S=\pm 1})=\delta_{l m} ,\quad
Tr({\cal P}_l^{S=\pm 1} \cdot {\cal L}_{m\,(tree)}^{S=\pm 1})=\delta_{l m}, 
\ee
where the trace is taken over spin and color indices, and 
${L}_{(tree)}^{S=\pm 1}$, ${\cal L}_{(tree)}^{S=\pm 1}$
are the tree-level amputated Green's functions of the operators 
${Q}^{S=\pm 1}$, ${\cal Q}^{S=\pm 1}$ respectively.

We impose the renormalization conditions:
\be
Tr(P_l^{S=\pm 1} \cdot {\hat L}_{m}^{S=\pm 1})=\delta_{l m} ,\quad
Tr({\cal P}_l^{S=\pm 1} \cdot \hat{\cal L}_{m}^{S=\pm 1})=\delta_{l m}.
\label{rencon} 
\ee
By inserting  Eqs. (\ref{rengreen}) in the above relations, we obtain
the renormalization matrices $Z^{S=\pm 1}$, ${\cal Z}^{S=\pm 1}$ in
terms of known quantities:
\be
Z^{S=\pm 1} = Z_{\Psi}^2 \left[ \left( D^{S=\pm 1} \right)^T \right]^{-1},\quad
{\cal Z}^{S=\pm 1} = Z_{\Psi}^2 \left[ \left( {\cal D}^{S=\pm 1} \right)^T \right]^{-1},
\ee 
where:
\be
D^{S=\pm 1}_{l m} \equiv Tr(P_l^{S=\pm 1} \cdot {L}_{m}^{S=\pm 1}) ,\quad
{\cal D}^{S=\pm 1}_{l m} \equiv Tr({\cal P}_l^{S=\pm 1} \cdot {\cal L}_{m}^{S=\pm 1}) . 
\ee

\noindent Note that $D^{S=\pm 1}$ and ${\cal D}^{S=\pm 1}$ have the same matrix
structure as $Z^{S=\pm 1}$ and ${\cal Z}^{S=\pm 1}$ respectively.

Due to lack of space we provide only the matrix ${\cal D}^{S=+1}$ 
(Parity Violating $P=-1$, Flavour exchange symmetry $S=+1$) for the special choices: 
$c_{SW}=0$, $\lambda=0$ (Landau Gauge), $r_s=r_d=r_{s'}=r_{d'}=1$, $N_c=3$, and
tree-level Symanzik action: 
%

%
{\small{
\be
{\cal D}^{S=+1}
=
\left(\begin{array}{rrrrr}
 {1+\cal D}_{11}  &0\hspace{0.55cm}         &0\hspace{0.55cm}         &0\hspace{0.55cm}        &0\hspace{0.55cm}  \\
 0\hspace{0.55cm}  &1+{\cal D}_{22}         &{\cal D}_{23}\hspace{0.25cm}  &0\hspace{0.55cm}        &0\hspace{0.55cm}  \\
 0\hspace{0.55cm}  &{\cal D}_{32}\hspace{0.25cm}  &1+{\cal D}_{33}     &0\hspace{0.55cm}        &0\hspace{0.55cm}  \\
 0\hspace{0.55cm}  &0\hspace{0.55cm}         &0\hspace{0.55cm}         &1+{\cal D}_{44}  &{\cal D}_{45}\hspace{0.25cm} \\
 0\hspace{0.55cm}  &0\hspace{0.55cm}         &0\hspace{0.55cm}         &{\cal D}_{54}\hspace{0.25cm}  &1+{\cal D}_{55}
\end{array}\right)^{S=+1}
\ee
}}
\hspace{-0.15cm}where:
{\small{
\bea
\Blue{{\cal D}_{11}}\hspace{-0.1cm}&=&\hspace{-0.1cm}+ \frac{g^2}{16\pi^2} \bigg[7.607190(2) + 2\,\ln(\Red{a^2} p^2) + \Big(2.642227(3) -\frac{19}{18}\,\ln(\Red{a^2} p^2)\Big)\,\Red{a^2} p^2  -2.79899088(3)\,\Red{a^2} \frac{\sum_\sigma p_\sigma^4}{p^2}\bigg],\nonumber \\
\Blue{{\cal D}_{22}}\hspace{-0.1cm}&=&\hspace{-0.1cm}+ \frac{g^2}{16\pi^2} \bigg[2.299519(2) + \ln(\Red{a^2} p^2) + \Big(1.846794(4) -\frac{25}{36}\,\ln(\Red{a^2} p^2)\Big)\,\Red{a^2} p^2  -0.87361421(2)\,\Red{a^2} \frac{\sum_\sigma p_\sigma^4}{p^2}\bigg],\nonumber \\
\Blue{{\cal D}_{23}}\hspace{-0.1cm}&=&\hspace{-0.1cm} - \frac{g^2}{16\pi^2} \bigg[1.1931473(4) + \Big(1.4685426(7) -\frac{1}{3}\,\ln(\Red{a^2} p^2)\Big)\,\Red{a^2} p^2  -0.89270364(3)\,\Red{a^2} \frac{\sum_\sigma p_\sigma^4}{p^2}\bigg],\nonumber \\
\Blue{{\cal D}_{32}}\hspace{-0.1cm}&=&\hspace{-0.1cm} - \frac{g^2}{16\pi^2} \bigg[10.970216(2) - 6\,\ln(\Red{a^2} p^2)+ \Big(6.711307(3) -\frac{7}{6}\,\ln(\Red{a^2} p^2)\Big)\,\Red{a^2} p^2  -1.7027590(1)\,\Red{a^2} \frac{\sum_\sigma p_\sigma^4}{p^2}\bigg],\nonumber \\
\Blue{{\cal D}_{33}}\hspace{-0.1cm}&=&\hspace{-0.1cm} + \frac{g^2}{16\pi^2} \bigg[11.595959(2) - 8\,\ln(\Red{a^2} p^2)+ \Big(3.102499(4) -\frac{4}{9}\,\ln(\Red{a^2} p^2)\Big)\,\Red{a^2} p^2  +1.92846914(2)\,\Red{a^2} \frac{\sum_\sigma p_\sigma^4}{p^2}\bigg],\nonumber \\
\Blue{{\cal D}_{44}}\hspace{-0.1cm}&=&\hspace{-0.1cm} + \frac{g^2}{16\pi^2} \bigg[10.269734(3) - 5\,\ln(\Red{a^2} p^2)- \Big(0.286209(5) -\frac{1}{18}\,\ln(\Red{a^2} p^2)\Big)\,\Red{a^2} p^2  +2.01567490(4)\,\Red{a^2} \frac{\sum_\sigma p_\sigma^4}{p^2}\bigg],\nonumber \\
\Blue{{\cal D}_{45}}\hspace{-0.1cm}&=&\hspace{-0.1cm} + \frac{g^2}{16\pi^2} \bigg[9.732710(2) - 5\,\ln(\Red{a^2} p^2)+ \Big(4.602710(4) -\frac{7}{9}\,\ln(\Red{a^2} p^2)\Big)\,\Red{a^2} p^2 -1.07855465(9)\,\Red{a^2} \frac{\sum_\sigma p_\sigma^4}{p^2}\bigg],\nonumber \\
\Blue{{\cal D}_{54}}\hspace{-0.1cm}&=&\hspace{-0.1cm} + \frac{g^2}{16\pi^2} \bigg[1.1783609(6) + \frac{1}{3}\,\ln(\Red{a^2} p^2)+ \Big(1.255191(1) -\frac{17}{54}\,\ln(\Red{a^2} p^2)\Big)\,\Red{a^2} p^2 -0.98220341(2)\,\Red{a^2} \frac{\sum_\sigma p_\sigma^4}{p^2}\bigg],\nonumber \\
\Blue{{\cal D}_{55}}\hspace{-0.1cm}&=&\hspace{-0.1cm} + \frac{g^2}{16\pi^2} \bigg[2.297078(2) + \frac{17}{3}\,\ln(\Red{a^2} p^2)+ \Big(0.828945(3) -\frac{23}{27}\,\ln(\Red{a^2} p^2)\Big)\,\Red{a^2} p^2 -3.06215785(3)\,\Red{a^2} \frac{\sum_\sigma p_\sigma^4}{p^2}\bigg].\nonumber
\eea
}}
%
\hspace{-0.15cm}In order to obtain $Z_{\Psi}$ for a given renormalization
prescription, one must make use of the inverse fermion propagator,
$S^{-1}$, calculated (up to 1-loop and up to ${\cal O}(a^2)$ for
massless Wilson/clover fermions and Symanzik gluons) in
Ref. \cite{Propolemiko}.


\end{document}